\documentclass[conference]{IEEEtran}
\IEEEoverridecommandlockouts
\usepackage{cite}
\usepackage{amsmath,amssymb,amsfonts}
\usepackage{algorithmic}
\usepackage{graphicx}
\usepackage{textcomp}
\usepackage{xcolor}
\usepackage{url}
\usepackage{tabularx}
\usepackage{subfig}
\usepackage{comment}

\def\tablename{Table}
\usepackage{etoolbox}
\makeatletter
\patchcmd{\@makecaption}
  {\scshape}
  {}
  {}
  {}
\makeatletter
\patchcmd{\@makecaption}
  {\\}
  {.\ }
  {}
  {}
\makeatother
\def\tablename{Table}
\def\BibTeX{{\rm B\kern-.05em{\sc i\kern-.025em b}\kern-.08em
    T\kern-.1667em\lower.7ex\hbox{E}\kern-.125emX}}
\begin{document}

\title{Scalable multi-chip quantum architectures enabled by 
cryogenic hybrid wireless/quantum-coherent network-in-package\\
}

 
\author{\IEEEauthorblockN{Eduard Alarc\'on\IEEEauthorrefmark{1},
 Sergi Abadal\IEEEauthorrefmark{1}, Fabio Sebastiano\IEEEauthorrefmark{2}, Masoud Babaie\IEEEauthorrefmark{2}, Edoardo Charbon\IEEEauthorrefmark{3}, Peter Haring Bol\'ivar\IEEEauthorrefmark{4}, \\
 Maurizio Palesi\IEEEauthorrefmark{5}, Elena Blokhina\IEEEauthorrefmark{6}, Dirk Leipold\IEEEauthorrefmark{6}, Bogdan Staszewski\IEEEauthorrefmark{7}, Artur Garcia-S\'aez\IEEEauthorrefmark{8} and Carmen G. Almudever\IEEEauthorrefmark{9}}

\IEEEauthorblockA{\IEEEauthorrefmark{1}\textit{Technical University of Catalunya, BarcelonaTech, Spain}}
\IEEEauthorblockA{\IEEEauthorrefmark{2}\textit{Delft University of Technology, The Netherlands}}
\IEEEauthorblockA{\IEEEauthorrefmark{3}\textit{École Polytechnique Fédérale de Lausanne, Switzerland}}
\IEEEauthorblockA{\IEEEauthorrefmark{4}\textit{University of Siegen, Germany}}
\IEEEauthorblockA{\IEEEauthorrefmark{5}\textit{University of Catania, Italy}}
\IEEEauthorblockA{\IEEEauthorrefmark{6}\textit{Equal 1, Ireland}}
\IEEEauthorblockA{\IEEEauthorrefmark{7}\textit{University College Dublin, Ireland}}
\IEEEauthorblockA{\IEEEauthorrefmark{8}\textit{Barcelona Supercomputing Center, Spain}}
\IEEEauthorblockA{\IEEEauthorrefmark{9}\textit{Technical University of Valencia, Spain}}}

\maketitle

\begin{abstract}

The grand challenge of scaling up quantum computers requires a full-stack architectural standpoint. In this position paper, we will present the vision of a new generation of scalable quantum computing architectures featuring distributed quantum cores (Qcores) interconnected via quantum-coherent qubit state transfer links and orchestrated via an integrated wireless interconnect.

\end{abstract}


\section{Proposed vision}

Today’s tremendous interdisciplinary efforts towards building a quantum computer is aimed at a machine capable of tackling problems beyond the reach of any classical computer. The so-called quantum advantage, a term referring to a quantum computer performing a specific computation that is intractable for a classical computer, has been recently claimed with state-of-the-art Noisy Intermediate-Scale Quantum (NISQ) computers consisting of several tens of quantum bits (qubits) \cite{arute2019quantum}. Nevertheless, it is widely recognized that addressing any real-world 
problem will require upscaling to thousands or even millions of qubits \cite{Preskill_2018}. Scaling quantum computers to such a large number of qubits is a major challenge due to, among others, the confluence of (i) technology factors confining the qubits to low fidelity, (ii) the need for cryogenic temperatures to reach practical coherence times, (iii) the dense integration of digital/RF control circuits, which are needed on a per-qubit basis \cite{staszewski2021cryo}, and (iv) the manifold architectural and algorithmic implications of managing noisy and short-lived 
qubits. The scaling-up race is fiercely complex and mostly revolves around the qubit technological aspects, but the bottleneck will very soon shift to the architectural problems stemming from the need to densely pack together the qubits and their classical electronics interfaces. 

\begin{figure}[t!]
	\centering
		\includegraphics[width=1\columnwidth]{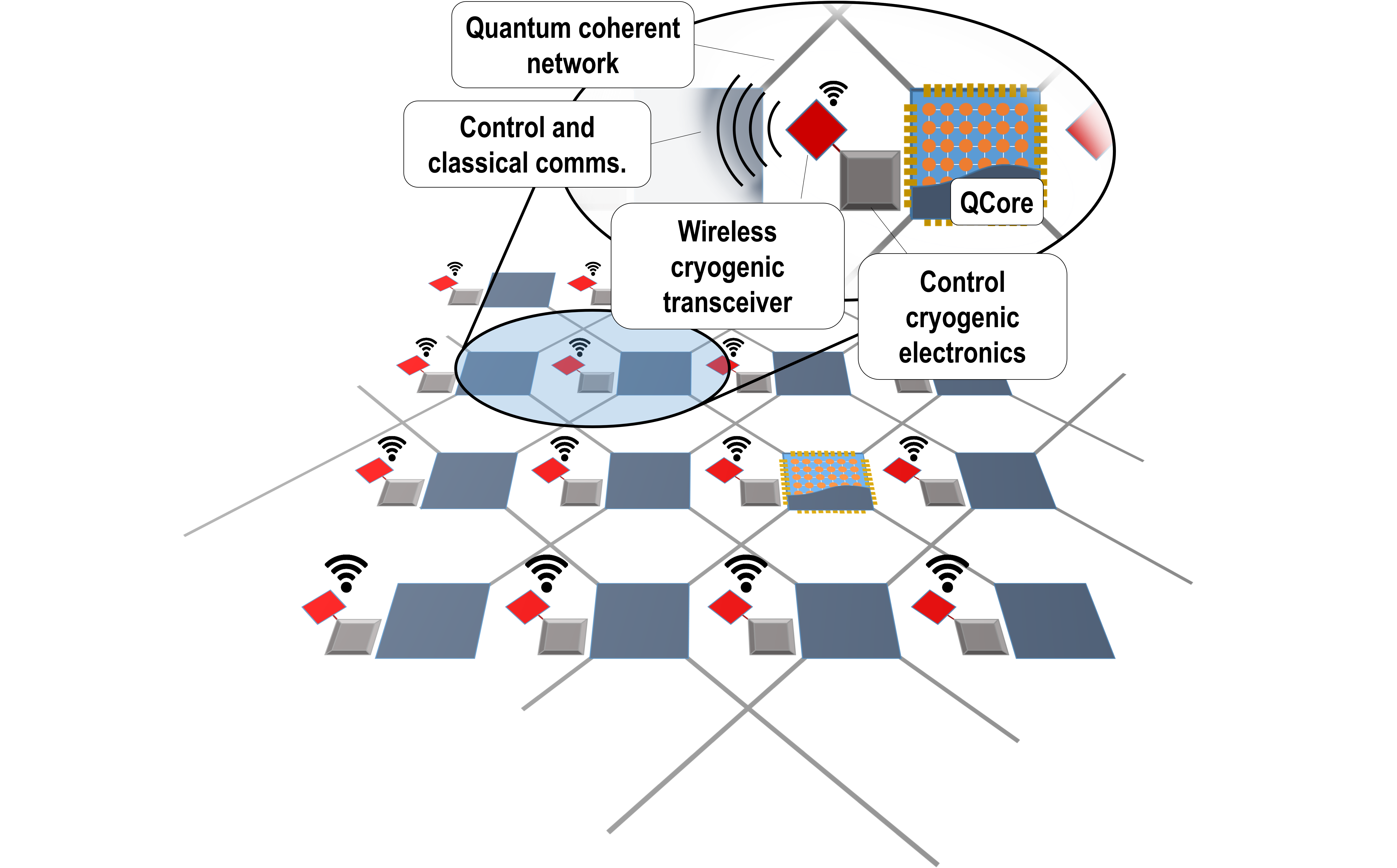}
 	\caption{The multi-Qcore architectural vision.}
 	\label{fig:vision}
\end{figure}

In analogy to early von Neumann computers, the first approach to the scaling of quantum computers has been to monolithically and densely integrate qubits within a single large array on the same silicon substrate, next to their classical control electronics. However, such a solution would impose an impractically high density of interconnects wiring the electronics at the array boundary to every single qubit or, in case of sharing of control lines, impractical requirements on qubit uniformity, increased noise (crosstalk) and limited yield. A viable architectural alternative proposed by leading quantum computing experts is to split the quantum processor into smaller cores (named here as Qcores) to be sparsely placed \cite{vandersypen2017interfacing, chow2021quantum, laracuente2022short}. This approach is by no means trivial, though, as it implies the development of (i) appropriate architectural means to seamlessly manage multiple quantum cores, i.e. a multi-Qcore architecture, (ii) an interconnect that supports the quantum and digital communication needs of such architecture, and (iii) a compact integrated quantum-coherent shared medium (i.e preserving quantum state coherence) to realize the quantum side of the interconnect. These key aspects are crucially missing and require a leap beyond the current technology to be realized.

In this context, this position paper envisions creating post-NISQ massive quantum processors through the development of scalable architectures with multiple Qcores interconnected via quantum links within the cryogenic package (see Figure \ref{fig:vision}). This is only possible by providing a quantum-coherent alternative to the rigid, static and wire-dense interconnects that are nowadays commonplace in quantum computers. Specifically, an ideal multi-Qcore interconnect fabric should not only support many simultaneous quantum state transfers across the Qcores, but also include a complementary control fabric capable of orchestrating such quantum state exchanges. If realized, this would unleash the architectural scalability potential of quantum computers by virtue of a balanced adaptive trade-off between communication and computation functions.

To realize this vision, this paper proposes an all-RF solution to the problem of building an integrated, scalable, and agile network spanning both quantum state and classical data transfers. This solution takes the form of multi-chip quantum cavity links compactly co-existing with wireless communication networks at cryogenic temperatures, both built on the solid grounds of cryogenic RF technology already developed for qubit interfacing. This approach encompasses solid experimental foundations of this interconnect fabric, but also demonstrating its paradigm-shifting architectural implications with a cross-cutting approach. In Section \ref{SecII}, this vision is further articulated, and in the subsequent sections the related multidisciplinary fields are discussed, with a final Section \ref{SecvII} devoted to a proposed design space exploration methodology across layers to assess the feasibility of this vision.

\section{Multi-core quantum computing architectures enabled by a hybrid interconnection network}

\label{SecII}


One of the main challenges for scaling up quantum computers is the wiring between the quantum processor and control electronics \cite{xue2021cmos}. Existing quantum computers physically span several temperature levels, with the qubits typically operating at sub-1-K temperatures to ensure their coherence, while the classical control electronics are operating at room temperature. While this thermal gap can be readily bridged by a few wires for the small quantum processors available today ($<$100 qubits), it would be impractical to wire thousands or millions of qubits in future quantum computers due to wiring size and reliability issues. Recent advances are closing this vertical gap by pushing the qubits to operate at temperatures near 4K \cite{petit2020universal}, where cryogenic refrigerators can dissipate the heat generated by the classical control electronics, and by developing cryogenic analog/RF control circuits that can be potentially co-integrated with the qubits at the same temperature level \cite{xue2021cmos}. Additionally, recent research advocates for also moving digital circuits to cryogenic temperatures \cite{schriek2020cryo}, an approach that further alleviates the wiring bottleneck. 


Although the full-cryogenic computer resulting from closing the vertical gap would be more compact than today’s prototypes, scaling issues remain in the horizontal dimension. Specifically, scaling this integrated scheme to a very large number of qubits as a single array (Qcore) is deemed unfeasible due to the practical bottleneck in the wiring between the qubits and their interface electronics \cite{vandersypen2017interfacing}. Moreover, other issues arise when integrating a high amount of qubits on a single chip, such as crosstalk, limited yield and qubit addressability. Instead, the multi-Qcore vision proposed here assumes $N_c$ quantum chips, with each chip hosting a Qcore composed of a moderately sized array of $N_q$ qubits. Thus, the Qcores can be spaced to fit the local electronics next to each core, thereby improving qubit addressability and reliability. Connecting the Qcores via quantum links allows the ensemble to remain coherent, thus maintaining the computational power of $N_c \times N_q$ entangled qubits, i.e. $O(2^{N_cN_q})$, rather than the incoherent addition of the Qcores, $O(N_c2^{N_q})$. The realization and integration of such quantum links at the chip scale is a fundamental challenge that has not been addressed yet.

The first key element to enable the vision is an expansive quantum channel as enabler of the quantum interconnect fabric. The other key element in the vision (in hybrid co-existence) is a multi-chip interconnection network that, with the exchange of classical data, coordinates the emitter and receiver of the quantum state transfer, and  enables the sharing of control signals and data across the Qcores. Without such an interconnection network, all data would have to go through the vertical connections to the host computer and back, which quickly becomes a roadblock. One could implement the inter-Qcore network through wired interconnects, yet this approach has drawbacks such as the difficulty of routing physical wires through the already densely wired system, or the high latency of system-wide transfers, which are critical for control. In this position paper, instead, we propose to realize the classical side of communications by means of a compact, high-bandwidth, highly reconfigurable wireless in-package network. The main idea is to integrate on-chip cryogenic antennas with RF cryogenic transceivers (developed with the same technology and techniques already employed in existing high-frequency qubit control circuits) so as to implement a wireless network within the quantum computing package. The resulting network is naturally system-wide and broadcast, allowing for agile reconfiguration of the underlying architecture. Moreover, although designing integrated circuits beyond their originally intended temperature operating range presents several challenges, the cryogenic environment is expected to opportunistically boost the antenna efficiency, reduce the thermal noise, and improve the transistor speed \cite{xue2021cmos,8036394}.

\section{Quantum-coherent high-Q planar cavity links}

Seamless micro-integration of cryogenic quantum communication links with quantum computational functions is the key enabler for the scaling and pervasive use of quantum information technology platforms \cite{awschalom2021development}. Optical photons have widely been used for high-fidelity remote entanglement and quantum state transfer \cite{ritter2012elementary}. However, optical quantum state transfer is highly probabilistic due to inefficiencies in photon coupling and transfer \cite{Monroe_2014}, placing fundamental limits on its use for inter-Qcore communication rates. In contrast, microwave cavities and circuits can combine low loss with strong coupling and are therefore well suited for on-demand high-rate quantum transfer, and thus to scale-up quantum computational architectures in a modular fashion. To date, a limited amount of microwave quantum communication solutions has been demonstrated, concentrating mainly in quantum register coupling \cite{axline2018demand}. Switchable quantum cavity channels to directly and resonantly entangle qubits from different Qcores are proposed to be implemented in the microwave and mm-wave frequency bands, and their CMOS-compatible heterogeneous integration into multi-Qcore solutions will be modeled, technological integration route developed, experimental demonstrator built-up and performance and limitations quantified. 

The qubits will be strongly coupled to a substrate dielectric waveguide \cite{giounanlis2019photon}, allowing direct, fast and tightly coupled transfer of quantum 
information from qubits to microwave photons and vice versa. Depending on the quality factor of the quantum cavity channel, the substrate material will be a hollow waveguide or sapphire which can provide low loss ($\tan \delta \thicksim 5\cdot 10^{-7}$) and 
quality factors exceeding $Q>10^6$. 
These features 
are essential for the efficient transmission and storage of microwave photons in the 
cavity. However, the high Q of the cavity leads to an extremely narrow-bandwidth frequency 
response, which requires tuning to and of the qubits’ resonance frequencies. This will be 
achieved by using a tunable coupler realized as a resonator with a tunable center frequency and tuning the qubit frequencies in resonance with the cavity waveguide. The planned dielectric waveguide can accommodate multiple modes, which can be used to selectively couple or decouple specific qubits in a multi-Qcore array, enabling frequency-multiplex quantum architectures. In this direction, two fundamental state-of-the-art challenges are addressed: (i) nanoscopic integration via plasmonic cavity approaches \cite{hosseininejad2017study} in order to provide a robust and monolithic integration technology with potential for tightly integrated wireless multi-Qcore communications, and (ii) low-loss, high quality-factor, low-noise coupling between neighboring Qcores providing inter-Qcore entanglement for massive quantum computation scalability. Microwave ($\thicksim$ 12 GHz) and mm-wave technologies are adressed to explore the advantages and limitations of alternative channel configurations.

\section{Cryo-CMOS RF transceivers}

Nanometer-scale CMOS electronics operating at cryogenic temperatures (cryo-CMOS) is widely accepted as the preferred IC technology for quantum-processor interfaces, thanks to its capability to operate down to (at least) 30 
mK, its unmatched very large scale of integration (VLSI), and the mature design automation infrastructure, all necessary to handle millions of qubits \cite{8036394, 8329135}. Several cryo-CMOS individual circuit blocks have been demonstrated, as required for qubit control and co-integration. A broad range of cryo-CMOS 
interface circuits for the control and readout of qubits has been recently shown, including complex microwave transmitters operating at 4 K 
providing $<$100 MHz data bandwidth/qubit for $<$-16 dBm output power 
at a carrier frequency $<$20 GHz \cite{xue2021cmos, 9209175}, RF receivers \cite{prabowo202113} and high-speed baseband data converters \cite{9365927}. Nevertheless, the lowest reported operating temperature for a wireless transceiver is 170 K \cite{4384375}, and no deep-cryogenic transceiver has been demonstrated. 

In this work, it is proposed to implement the first-ever deep-cryogenic wireless transceiver, operating down to 4 K. To miniaturize the on-chip antenna, the operating frequency is expected to increase $>3\times$ beyond the state-of-the-art for cryo-CMOS, while 
the data bandwidth and the transmitter’s output power will both be extended by $>20\times$
beyond the state of the art to realize a robust 
communication between the qubit tiles. We will explore using the waveguide proposed in section II for classical data transmission at 60 GHz by exciting the waveguide with an on-chip antenna and by minimizing the crosstalk to the quantum  channels. 

\section{Communication Intranet within a Quantum Computer}

For an appropriate architectural integration of all-RF interconnects, the quantum photon and RF propagation within a quantum package needs to be characterized, and adequate communication 
protocols must be developed. On the one hand, existing works in channel modeling are mostly simulation-based studies for conventional computing chips at room temperature, at the mm-wave band and in the frequency domain only~\cite{abadal_ieeeaccess20}. In contrast, for quantum computers at cryogenic temperatures, this field is at its infancy and the only relevant work is at microwave frequencies and not oriented to communications~\cite{huang_prxquantum21}. Hence, there are no models capable of capturing the particularities of the channels within quantum computer packages and cavities, which differ substantially from those in conventional computers, and at cryogenic temperature. We propose to explore 
this uncharted territory and provide a complete channel model in the time and frequency domains. 

Since it appears to be the first time that multi-chip wireless/quantum interconnects are proposed for quantum computers, adequate protocols are missing. Protocols for quantum communications are currently limited 
to large-scale Quantum Internet proposals, which are grounded on the distribution of entangled pairs with repeaters at kilometer distances~\cite{dahlberg_sigdc19}. These assumptions do not apply to the proposed vision scenario, which is based on cavity-enabled quantum coupling at chip-scale distance and shows tight interplay with computation and high latency criticality. On the wireless transmission of classical data at the chip scale, the existing research has focused on theoretical wireless on-chip networks for conventional processors with tens of transmitters. MAC protocols are 
generally variants of token passing, seeking simplicity and performance~\cite{abadal_ieeecm18}; whereas, at the network layer, wireless links are generally considered fixed unicast connections between distant cores~\cite{mansoor_tmscs15}. None of these works has therefore considered the unique problem addressed in this position paper, namely the need for protocols to manage the access to potentially hundreds of latency-critical, computing-driven quantum links at cryogenic temperatures, with the implications that this has on the design of protocols for the non-quantum wireless links. We propose to develop a protocol stack covering both the quantum and non-quantum 
planes and capable of systematically leveraging architectural information to maximize the system performance.

\section{Architecting Scalable and Reconfigurable Quantum Processors}
NISQ devices are publicly accessible (e.g., IBM Q experience, Qinspire) and users can already run small instances of quantum algorithms. However, these quantum processors suffer from several constraints, such as limited connectivity 
among the qubits. This requires the quantum compiler to modify the quantum algorithm, described as a quantum circuit, to realize it on a given quantum chip. 
This transformation process is known as mapping and might compromise the successful execution of the algorithm \cite{almudever2020realizing}. 

Exact approaches for small-size 
quantum circuits and approximate mapping solutions using heuristics for larger quantum circuits have been developed for specific single-core NISQ 
devices \cite{lao2021timing,murali2019full}. Recently, the first compiler techniques for mapping and scheduling quantum algorithms onto connectivity-simplified multi-core quantum 
architectures have been proposed \cite{baker2020time,rodrigo2021double} as distributed or modular architectural approaches are becoming more prominent for scaling-up quantum hardware \cite{vandersypen2017interfacing, chow2021quantum, laracuente2022short,brown2016co}. These 
mapping solutions all focus on multi-Qcore architectures that assume all-to-all intra- and inter-core connectivity with unlimited and ideal communication resources and a fixed and invariant interconnection network. In this work it is proposed to
build on top of these quantum hardware advances and define scalable multi-Qcore architectures, including the communication perspective and the required compilation techniques, connecting today's experiments with future practical implementations. We will also perform on-line dynamic optimizations of the Qcores transfers for an efficient execution of the algorithms leveraging the reconfigurability provided by the wireless control plane.

\section{Design Space Exploration for Architecture/Network/Circuit/link Co-Design}

\label{SecvII}

Although quantum computers already exist in the form of intermediate-scale quantum processors, there is no consensus in the quantum computing community on what benchmarks and metrics to use to compare them and to assess their performance. As a first attempt, IBM proposed Quantum Volume (QV) and Circuit Layer Operations Per Second (CLOPS) as metrics to measure three key quantum computing performance attributes: quality, speed and scale \cite{wack2021quality}. Also, different sets of quantum benchmarks for assessing their overall performance have been introduced \cite{blume2020volumetric, mills2021application}. In addition, the higher layers of full-stack systems, consisting of both quantum software and classical control hardware, have been so far designed and provided solutions following a bottom-up approach; that is, they have been developed for a specific quantum device. This approach is appropriate for a single rigid design, which could be functional, but it is unclear how close it is to the optimal design in terms of overall performance and scalability, and how such performance holds in different specifications. This is particularly critical in the harsh constraints of quantum computing. In addition, a long-standing perceived need from the quantum community identifies the lack of design guidelines from top architectural layers down to physical ones or even cross-layer co-design \cite{almudever2021structured, tomesh2021quantum}. 

\begin{figure}[t!]
	\centering
	\includegraphics[width=0.8\linewidth]{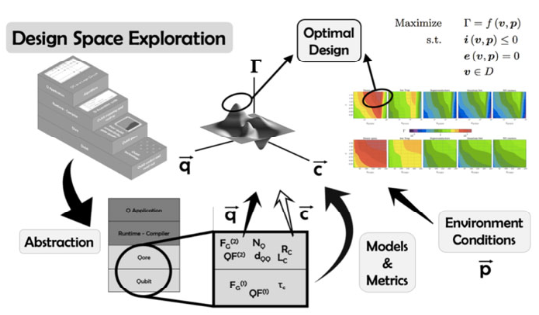}
 	\caption{Exploration framework for quantum computing architectures.}
 	\label{fig:DSE}
\end{figure}

In this work, we propose to employ structured Design Space Exploration (DSE) methodologies to perform a cross-layer co-design of the full-stack quantum system, including both communication and computation, allowing for both top-down and bottom-up optimizations across layers (see Figure \ref{fig:DSE}). To this purpose, models will be developed that describe (i) the computational part, from the distributed architecture down to the single Qcore and qubit impairments, and (ii) the classical and quantum communication parts from the perspective of performance (e.g. throughput, fidelity) and efficiency (e.g. power). Finally, we will create and profile a set of large-scale benchmarks able to stress the multi-Qcore platform and define different performance metrics of the overall architecture.

\section{Conclusions}
This position paper addresses from a full-stack architectural perspective the challenge of scaling up quantum computers.  A vision is presented of a new generation of scalable quantum computing architectures in which distributed quantum cores consider quantum-coherent qubit state transfer links in the data plane to interconnect, and a wireless medium orchestrator. The scientific and technology needs at each architectural layer are critically discussed, and a cross-layer design-space exploration framework is proposed to assess the feasibility of the proposed architectures.

\section*{Acknowledgments}
All authors acknowledge support from the EU, grant HORIZON-ERC-101042080 and grant HORIZON-EIC-2022-PATHFINDEROPEN-01-101099697, QUADRATURE.

\end{document}